\documentclass{Interspeech}
\usepackage[T1]{fontenc}

\interspeechcameraready 

\title{EmoSSLSphere: Multilingual Emotional Speech Synthesis with Spherical Vectors and Discrete Speech Tokens}

\def\anonname{Joonyong Park$^{1,2}$, Kenichi Nakamura$^2$}
\def\anonaddress{
  $^1$The University of Tokyo, Japan,
  $^2$Preferred Networks Inc., Japan
}
\def\anonemail{jpark@gavo.t.u-tokyo.ac.jp, knakamura@preferred.jp}

\keywords{Emotional Speech Synthesis, Multilingual Speech Synthesis, Spherical Emotion Vectors, Self-Supervised Learning (SSL)}

\begin{document}

\maketitle

\begin{abstract}
This paper introduces EmoSSLSphere, a novel framework for multilingual emotional text-to-speech (TTS) synthesis that combines spherical emotion vectors with discrete token features derived from self-supervised learning (SSL). By encoding emotions in a continuous spherical coordinate space and leveraging SSL-based representations for semantic and acoustic modeling, EmoSSLSphere enables fine-grained emotional control, effective cross-lingual emotion transfer, and robust preservation of speaker identity.

We evaluate EmoSSLSphere on English and Japanese corpora, demonstrating significant improvements in speech intelligibility, spectral fidelity, prosodic consistency, and overall synthesis quality. Subjective evaluations further confirm that our method outperforms baseline models in terms of naturalness and emotional expressiveness, underscoring its potential as a scalable solution for multilingual emotional TTS.

\end{abstract}
\section{Introduction}

Recent advances in deep learning have significantly enhanced text-to-speech (TTS) synthesis, enabling the generation of speech that closely resembles natural human voices~\cite{ju2024naturalspeech3zeroshotspeech, chen2024valle2neuralcodec}. However, multilingual emotional speech synthesis remains challenging due to difficulties in maintaining speaker individuality, consistent emotional expression across languages, and natural prosody~\cite{kang2023zetspeechzeroshotadaptiveemotioncontrollable, liu2023promptstylecontrollablestyletransfer, wang2023learningemotionalrepresentationsimbalanced}. Existing approaches relying on explicit emotion embeddings or cross-lingual transfer models often produce unnatural intonation, speaker drift, and inconsistent emotional delivery when applied across languages.

A notable recent approach, EmoSphere-TTS~\cite{Cho_2024}, demonstrated the effectiveness of using spherical representations of Arousal–Valence–Dominance (AVD) emotional vectors, enabling smooth and intuitive emotional control. Nonetheless, EmoSphere-TTS has limited capacity in modeling fine-grained prosodic nuances and ensuring linguistic fidelity, particularly in multilingual contexts.

In this study, we propose \textbf{EmoSSLSphere}, a multilingual emotional TTS framework designed to overcome these limitations. EmoSSLSphere integrates spherical emotion vectors with discrete speech tokens learned via self-supervised learning (SSL), enabling precise emotional control and consistent prosodic patterns across multiple languages. Specifically, SSL-derived token representations capture language-agnostic prosodic features, while spherical emotion vectors offer an interpretable continuous emotional space. By combining these components, EmoSSLSphere achieves enhanced multilingual emotional expressiveness, naturalness, and speaker individuality in synthesized speech. Our audio samples are available at \texttt{nonmetal.github.io/emossl-page-demo}. 

\vspace{-2mm}
\section{Prior Studies}

Emotionally expressive TTS systems typically rely on explicit emotion labels as additional inputs to guide the generation of speech with diverse affective styles~\cite{kharitonov2022textfreeprosodyawaregenerativespoken, ma2023emotion2vecselfsupervisedpretrainingspeech, zhou2024emotionaldimensioncontrollanguage}. While these label-based methods have shown effectiveness in controlled settings, they often struggle to model nuanced emotional expressions and smooth transitions between different emotional states~\cite{wang2023emotioncontrollabletts}.

To address these limitations, recent approaches have introduced methods that jointly encode emotional and speaker-related attributes. For instance, iEmoTTS~\cite{Zhang2023iEmoTTS} explicitly disentangles prosody and timbre using supervised style encoders to improve speaker consistency in emotion transfer. Similarly, multi-layered emotion modeling strategies~\cite{shi2024mmmmultilayermultiresidualmultistream} attempt to separate language-independent emotional features from language-specific attributes. Despite these advances, these methods generally require explicit supervision and still fall short of providing intuitive and precise emotional controllability.

To enhance controllability and interpretability in emotional modeling, Arousal–Valence–Dominance (AVD)-based representations have emerged as a promising continuous emotion modeling approach~\cite{Habib2020Semi-Supervised, sivaprasad21_interspeech}. In particular, EmoSphere-TTS~\cite{Cho_2024} introduces spherical transformations of AVD vectors, facilitating smoother and more intuitive emotional transitions. Building upon this, EmoSphere++~\cite{cho2024_pp} further extends the framework through semi-supervised training methods, improving data efficiency and reducing the reliance on large annotated emotional datasets. However, both EmoSphere-TTS and EmoSphere++ primarily focus on monolingual scenarios and do not address the complexities involved in multilingual emotional synthesis.

Multilingual emotional TTS presents additional challenges due to phonological, prosodic, and emotional variability across languages. To overcome these challenges, researchers have proposed cross-lingual emotion embeddings to transfer emotional expressions across languages~\cite{zhu2023mettsmultilingualemotionaltexttospeech}. Nevertheless, these methods frequently suffer from unnatural prosody, foreign accents, or emotional drift. Additionally, disentanglement-based approaches such as iEmoTTS remain challenging to scale effectively in multilingual and low-resource scenarios.

More recently, zero-shot multilingual emotional synthesis frameworks utilizing self-supervised learning (SSL) have been explored to achieve language and speaker adaptation without extensive annotated data~\cite{saeki23_inproceedings,10.1109/TASLP.2024.3451951}. However, such frameworks often exhibit limited control and consistency in emotional expression across different languages. Other techniques employing vector quantization (VQ)-based methods~\cite{Chen2023AVQ, 9747098} or perturbation-based regularization~\cite{zhou2022mix, Zhou_2023} have shown improvements in prosodic alignment and reduced speaker interference, but these approaches have not yet thoroughly explored their applicability to multilingual consistency.

\begin{figure}[tb]
    \centering
    \includegraphics[width=7.9cm]{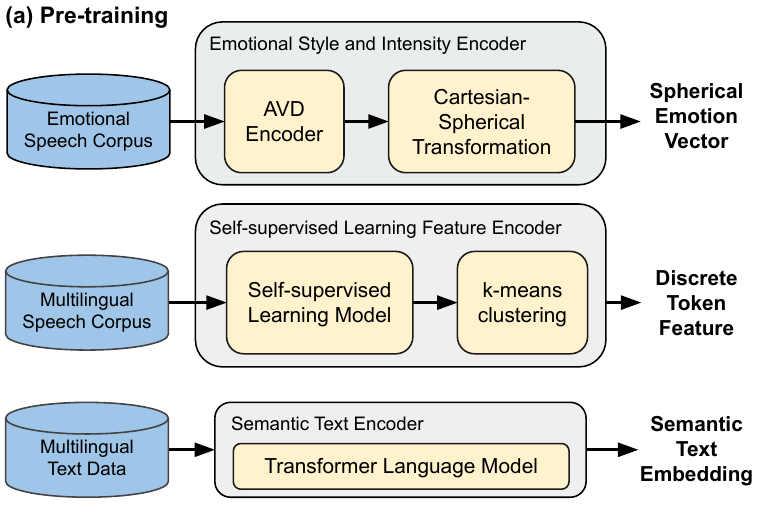}
    \caption{Proposed model architecture in (a) pre-training stage.}
    \label{fig:fig/pre_training.pdf}
    \vspace{-2mm}
\end{figure}

\begin{figure*}
    \centering
    \includegraphics[width=14.9cm]{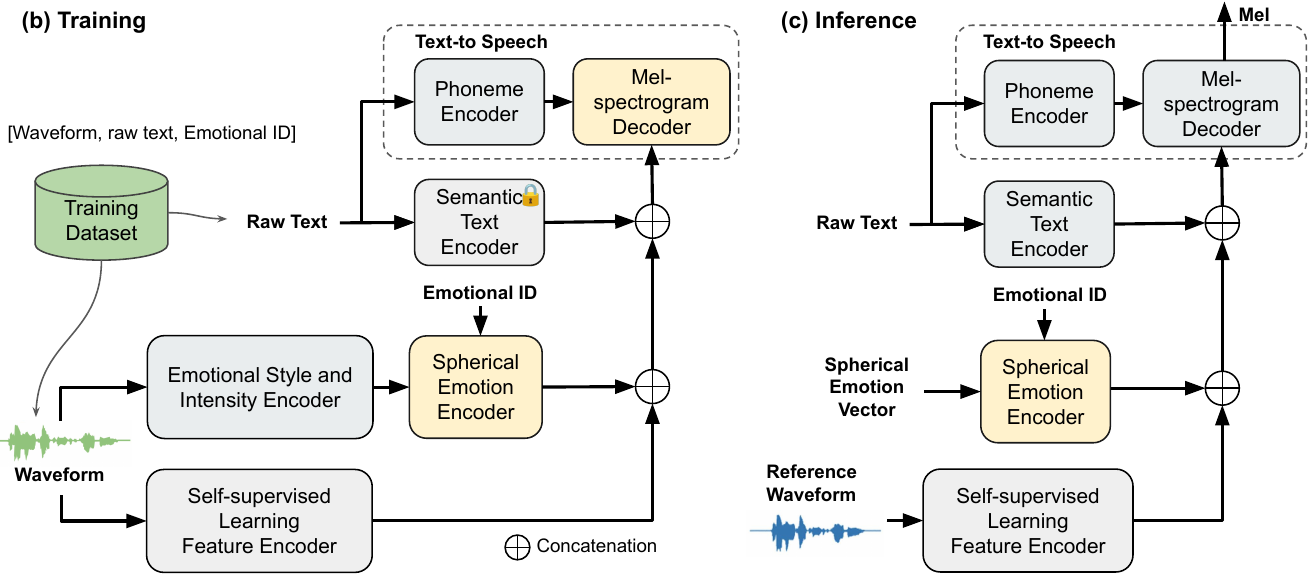}
    \caption{Proposed model architecture in (b) training and (c) inference stage.}
    \label{fig:fig/training_inference.pdf}
    \vspace{-2mm}
\end{figure*}

\section{Model Architecture}

Although existing methods demonstrate notable progress in emotional speech synthesis, challenges remain in achieving effective integration of linguistic, emotional, and acoustic features, especially in multilingual settings. Current models often struggle with accurately preserving semantic fidelity, speaker individuality, and fine-grained emotional controllability simultaneously. To address these issues, we propose \textbf{EmoSSLSphere}, a multilingual emotional TTS framework that leverages spherical emotional vectors and self-supervised learning (SSL)-derived discrete token representations. EmoSSLSphere extends the EmoSphere-TTS~\cite{Cho_2024} architecture, enhancing its capability for multilingual adaptation, precise emotional control, and robust prosodic modeling.

The EmoSSLSphere architecture consists of four primary modules: an \textbf{Emotional Style and Intensity Encoder}, an \textbf{SSL-based Feature Encoder}, a \textbf{Text Encoder}, and a \textbf{Mel-spectrogram Decoder}. Each encoder independently captures complementary speech characteristics, and their embeddings are fused to provide rich input for the TTS decoder.

\subsection{Emotional Style and Intensity Encoder}
Following the EmoSphere-TTS approach~\cite{Cho_2024}, we adopt a continuous emotional representation based on the Arousal–Valence–Dominance (AVD) model~\cite{Habib2020Semi-Supervised, sivaprasad21_interspeech}. AVD vectors are extracted from the Mel-spectrograms of reference speech using a pretrained encoder, and subsequently transformed from Cartesian $(x, y, z)$ coordinates into spherical $(r, \theta, \phi)$ coordinates. This transformation facilitates smooth interpolation across emotional states and enables fine-grained, continuous control over emotional intensity.

In addition to the continuous AVD representation, we incorporate discrete Emotion IDs, which are represented as one-hot vectors and embedded into a categorical style vector. These vectors represent prototypical emotion classes (e.g., \textit{Angry}, \textit{Sad}, \textit{Happy}), and complement the spherical AVD by explicitly encoding stylistic priors tied to these classes. The concatenation of these two types of emotional information allows the model to jointly benefit from both categorical emotion specificity and continuous emotional nuance.

During inference, both the spherical AVD vector and the embedded Emotion ID are derived from reference audio and fused to guide the prosody and affective expressiveness of the generated speech. This hybrid formulation was empirically found to improve robustness and expressiveness, especially in low-resource or ambiguous emotion scenarios.

\subsection{SSL-based Feature Encoder}
To model language-independent prosodic cues, we adopt HuBERT~\cite{hsu2021hubert}, a self-supervised model trained on large-scale speech corpora.  
Following prior analyses that show lower–middle transformer layers encode rhythm and pitch trajectories reliably across languages~\cite{hsu2021hubert}, we extract the hidden states from the \textbf{9th layer} in 12-layer base model, as it provides a favorable trade-off between fine phonetic detail (lower layers) and high-level semantics (upper layers).

The continuous dimensional features are then \emph{vector-quantized} via $k$-means clusterization. Clustering is performed separately for each language so that phone-inventory differences do not blur the codebook; however, the resulting tokens were found to capture universal prosodic patterns are weakly language specific. These discrete prosody tokens provide a stable, interpretable control signal for rhythm, pitch contour, and intensity, and—in conjunction with the continuous AVD representation, enabling the decoder to realise expressive speech while preserving the intended emotional trajectory.

\subsection{Text Encoder} 
The Text Encoder consists of two parallel components: a \textit{Phoneme Encoder} and a \textit{Semantic Encoder}. The phoneme encoder processes input text into phoneme sequences using forced alignment tools, ensuring accurate pronunciation and intelligibility. This phoneme-level representation is the primary linguistic input to the decoder.

In parallel, the semantic encoder leverages DeBERTaV3~\cite{debertav3}, a transformer-based language model trained on multilingual corpora, to produce higher-level semantic embeddings. These semantic embeddings are first used to condition the emotional and prosodic modules through cross-attention mechanisms. After linear projection, they are also concatenated with the phoneme, emotion, token, and speaker embeddings as part of the decoder input, as shown in Figure~2. To ensure multilingual applicability, only the tokenizer is language-adapted while the core encoder is kept frozen.

\subsection{Training Procedure} Training consists of two distinct phases: \textit{pre-training} of individual encoders and \textit{fine-tuning} of the decoder.

\begin{itemize} 
    \item[1.] \textbf{Pre-training Stage:} The Emotional Style and Intensity Encoder is pre-trained following EmoSphere-TTS, converting Mel-spectrogram-derived AVD values into spherical emotional embeddings. Independently, the SSL-based Feature Encoder is pre-trained using HuBERT features discretized by language-specific k-means clustering. Finally, the Semantic Text Encoder utilizes pre-trained multilingual DeBERTaV3 embeddings without further fine-tuning. (Figure 1)
    
    \item[2.] \textbf{Fine-tuning Stage:}
    In this stage, the Mel-spectrogram Decoder (based on FastSpeech2~\cite{fs2cite}) is trained using paired text-waveform data. All encoder parameters are kept frozen during decoder training. Input embeddings (phoneme sequences, spherical emotion vectors, discrete tokens, speaker embeddings, and semantic embeddings) are first projected into a shared latent space, concatenated, and then linearly projected to match the decoder input dimension. This ensures effective fusion and conditioning across linguistic, prosodic, emotional, and speaker-related dimensions. (Figure 2, left)
\end{itemize}

\subsection{Inference Procedure} 
During inference, EmoSSLSphere requires two inputs: the target text and a reference speech waveform. The reference waveform provides emotional and prosodic cues and is typically sourced from the same speaker or speaker cluster as the target output. It does not need to match the text content but should reflect the desired emotional tone and prosodic style.

The emotional embeddings and discrete tokens are extracted from this reference and used to condition the generated speech's intensity, style, and prosody. Meanwhile, the phoneme encoder processes the target text for pronunciation and intelligibility. Speaker identity is preserved by using a fixed speaker embedding corresponding to the target voice, independently of the reference audio. The combined embedding vector is passed to the Mel-spectrogram decoder, which generates expressive and speaker-consistent synthesized speech. (Figure 2, right)

\begin{figure}[tb]
    \centering
    \includegraphics[width=7.9cm]{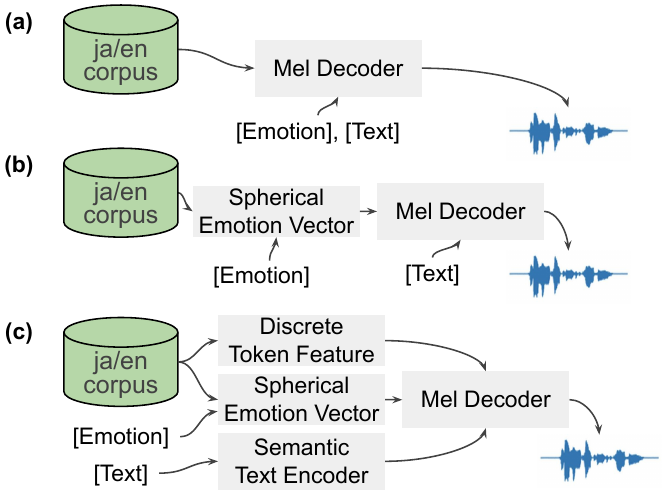}
    \caption{Architectures of conventional models: (a) NATSpeech with Emotion Label and (b) EmoSphereTTS, which are compared with the proposed method (c) EmoSSLSphere for evaluating speech synthesis systems.}
    \label{fig:fig/analysis_method.pdf}
    \vspace{-2mm}
\end{figure}

\begin{table*}[h]
    \centering
    \caption{\textit{Evaluation results on speech intelligibility. The last group of rows represents an ablation study of the proposed model.}}
    \resizebox{\linewidth}{!}
    {
    \begin{tabular}{lcccc|cccc}
        \toprule
         & \multicolumn{4}{c}{\textbf{English}} & \multicolumn{4}{c}{\textbf{Japanese}} \\
        \toprule
        \textbf{Method} & \textbf{WER (\%)} $\downarrow$ & \textbf{SpeechBERTScore} $\uparrow$ & \textbf{SpeechBLEU} $\uparrow$ & \textbf{SpeechTokenDist.} $\uparrow$ & \textbf{CER (\%)} $\downarrow$ & \textbf{SpeechBERTScore} $\uparrow$ & \textbf{SpeechBLEU} $\uparrow$ & \textbf{SpeechTokenDist.} $\uparrow$ \\
        \midrule
        GT & \textbf{13.65} & - & - & - & \textbf{16.37} & - & - & - \\
        \midrule
        NATSpeech w/ Emotion Label \cite{ren2022portaspeechportablehighqualitygenerative}  & 21.50 & 0.870 & 0.200 & 0.585 & 20.30 & 0.869 & 0.180 & 0.588   \\
        EmoSphereTTS  \cite{Cho_2024} & 20.96 & 0.875 & 0.209 & 0.591 & 19.26 & \textbf{0.874} & 0.188 & 0.593  \\
        \midrule
        \textbf{EmoSSLSphere (Proposed)} &  \textbf{19.58} & 0.880 & \textbf{0.225} & \textbf{0.602} & \textbf{18.33} & 0.873 & \textbf{0.218} & 0.594   \\
        \quad w/o Reference wavefile & 20.37 & 0.874 & 0.212 & 0.595 & 19.18 & 0.870 & 0.200 & 0.590  \\
        \quad Ref. wav (Content unmatch) & 20.31 & 0.873 & 0.210 & 0.592 & 19.11 & 0.866 & 0.202 & 0.589 \\
        \quad Ref. wav (Emotion unmatch) & 19.83 & 0.875 & 0.219 & 0.598 & 18.79 & 0.871 & 0.212 & 0.590 \\
        \quad Ref. wav (Speaker unmatch)& 19.69 & \textbf{0.881} & 0.224 & 0.602 & 18.42 & 0.871 & 0.215 & \textbf{0.596} \\
        \quad w/o Semantic Text Encoder & 19.72 & 0.879 & 0.223 & 0.600 & 18.46 & 0.872 & 0.216 & 0.593  \\
        \quad w/o k-means discretization & 20.12 & 0.876 & 0.217 & 0.597 & 18.89 & 0.871 & 0.208 & 0.591  \\
        \bottomrule
    \end{tabular}
    }
\end{table*}

\begin{table*}[h]
    \centering
    \caption{\textit{Evaluation results on speech quality and naturalness. The last group of rows represents an ablation study of the proposed model. The nMOS scores are presented with 95\% confidence intervals.}}
    \resizebox{\linewidth}{!}
    {
    \begin{tabular}{lccccc|ccccc}
        \toprule
         & \multicolumn{5}{c}{\textbf{English}} & \multicolumn{5}{c}{\textbf{Japanese}} \\
        \toprule
        \textbf{Method} & \textbf{MCD} $\downarrow$ & \textbf{LogF0RMSE} $\downarrow$ & \textbf{WARP-Q}$\uparrow$ & \textbf{UTMOSv2} $\uparrow$ & \textbf{nMOS} $\uparrow$ & \textbf{MCD} $\downarrow$ & \textbf{LogF0RMSE} $\downarrow$ & \textbf{WARP-Q}$\uparrow$ & \textbf{UTMOSv2} $\uparrow$ & \textbf{nMOS} $\uparrow$\\
        \midrule
        GT & - & - & - & - & \textbf{4.30$\pm$0.09} & - & - & - & - & \textbf{4.40$\pm$0.12}  \\
        \midrule
        NATSpeech w/ Emotion Label \cite{ren2022portaspeechportablehighqualitygenerative} & 8.060 & 0.357 & 3.413 & 2.348 & - & 9.287 & 0.433 & 3.736 & 1.790 & -  \\
        EmoSphereTTS  \cite{Cho_2024} & 7.979 & 0.359 & 3.400 & 2.403 & 4.05$\pm$0.12 & 9.131 & 0.428 & 3.724 & 1.875 & 3.63$\pm$0.16  \\
        \midrule
        \textbf{EmoSSLSphere (Proposed)} & \textbf{7.282} & \textbf{0.341} & \textbf{3.465} & 2.427 & \textbf{4.13$\pm$0.08} & \textbf{8.719} & \textbf{0.409} & \textbf{3.744} & 1.912 & \textbf{3.94$\pm$0.15}  \\
        \quad w/o Reference wavefile & 7.732 & 0.356 & 3.411 & \textbf{2.513} & - & 9.134 & 0.425 & 3.734 & 1.788 & -  \\
        \quad Ref. wav (Content unmatch) & 7.681 & 0.354 & 3.438 & 2.502 & - & 9.081 & 0.423 & 3.730 & 1.804 & - \\
        \quad Ref. wav (Emotion unmatch) & 7.405 & 0.349 & 3.433 & 2.487 & - & 8.997 & 0.420 & 3.733 & 1.819 & - \\
        \quad Ref. wav (Speaker unmatch)& 7.523 & 0.345 & 3.439 & 2.474 & - & 8.846 & 0.417 & 3.738 & 1.835 & - \\
        \quad w/o Semantic Text Encoder & 7.293 & 0.349 & 3.460 & 2.479 & - & 8.809 & 0.413 & 3.744 & \textbf{1.980} & -  \\
        \quad w/o k-means discretization & 7.418 & 0.348 & 3.436 & 2.441 & -  & 8.812 & 0.419 & 3.722 & 1.845 & - \\
        \bottomrule
    \end{tabular}
    }
\end{table*}

\begin{table*}[h]
    \centering
    \caption{\textit{Evaluation results on emotional expressiveness, AVD RMSE per emotion.}}
    \resizebox{\linewidth}{!}
    {
    \begin{tabular}{lccccc|ccccccc}
        \toprule
         & \multicolumn{5}{c|}{\textbf{English}} & \multicolumn{7}{c}{\textbf{Japanese}} \\
        \toprule
        \textbf{Method} & \textbf{Avg.} $\downarrow$ & Angry & Happy  & Sad  & Surprise  & \textbf{Avg.} $\downarrow$ & Angry & Happy  & Sad  & Surprise & Disgust & Fear \\
        \midrule
        NATSpeech w/ Emotion Label \cite{ren2022portaspeechportablehighqualitygenerative} 
        & 0.0938 & \textbf{0.0736} & 0.1221 & 0.0908 & 0.0885 
        & 0.0936 & \textbf{0.0558} & 0.0851 & 0.1474 & 0.0801 & 0.1212 & 0.0719 \\
        
        EmoSphereTTS \cite{Cho_2024} 
        & 0.0798 & 0.0663 & 0.0870 & 0.0913 & \textbf{0.0747} 
        & 0.0867 & 0.0822 & \textbf{0.0742} & \textbf{0.1159} & 0.0791 & 0.1024 & 0.0665 \\

        \midrule
        \textbf{EmoSSLSphere (Proposed)}  
        & \textbf{0.0773} & 0.0837 & 0.0658 & \textbf{0.0757} & 0.0839 
        & \textbf{0.0783} & 0.0601 & 0.0544 & 0.1488 & \textbf{0.0610} & \textbf{0.0854} & \textbf{0.0600} \\
        \bottomrule
    \end{tabular}
    }
\end{table*}
\vspace{-1mm}

\section{Experimental Setup}

\subsection{Model Configuration}

Our proposed model, EmoSSLSphere, builds upon the NATSpeech architecture~\cite{ren2022portaspeechportablehighqualitygenerative} and incorporates additional modules to support multilingual and emotionally expressive speech synthesis, as shown in Figure 3. The architecture comprises multiple encoder modules and a decoder trained in a two-stage setup, where encoders are frozen during decoder training.

To accommodate phonological and semantic differences across languages, we instantiate separate encoder modules (Emotional Encoder, SSL Feature Encoder, and Text Encoder) for each target language, thereby avoiding cross-lingual interference and allowing language-specific adaptation.

\begin{itemize} 
    \item \textbf{Emotional Style and Intensity Encoder:}
    Following EmoSphere-TTS~\cite{Cho_2024}, we extract AVD (Arousal-Valence-Dominance) vectors from reference Mel-spectrograms using a pre-trained encoder. These vectors are transformed from Cartesian to spherical coordinates, enabling smooth and continuous emotional control. We also incorporate categorical Emotion ID embeddings, which are concatenated with the spherical AVD vectors to capture both global emotional style and discrete emotion categories.

    \item \textbf{SSL-based Feature Encoder:} 
    To capture prosodic structure and acoustic style, we utilize the HuBERT-base model~\cite{hsu2021hubert}, pre-trained on LibriSpeech~\cite{librispeech}. Latent representations from the 9th transformer layer are extracted and discretized via k-means clustering ($K=200$), trained independently on emotion-rich English and Japanese datasets~\cite{zhou2022emotionalvoiceconversiontheory, jvnv}. These tokens represent segmental prosodic cues such as pitch, rhythm, and intensity, which support expressive synthesis in a speaker-agnostic fashion.

    \item \textbf{Text Encoder:} 
    The text encoder consists of two parallel components: a phoneme encoder and a semantic encoder. For phoneme encoding, we use Montreal Forced Alignment (MFA) for English and pyopenjtalk\footnote{https://github.com/r9y9/pyopenjtalk} from ESPNet~\cite{1804.00015} for Japanese. This phoneme-level output is directly passed to the decoder. In parallel, the semantic encoder uses DeBERTaV3~\cite{debertav3}, with the base model for English and a Japanese version fine-tuned on Wikipedia and Aozora Bunko Corpus\footnote{https://www.kaggle.com/datasets/ryancahildebrandt/azbcorpus}. The semantic output is not directly input to the decoder but instead conditions the emotional and prosodic encoders to align speech with the intended meaning.

    \item \textbf{Decoder:} 
    The decoder, based on FastSpeech2~\cite{fs2cite}, generates the Mel-spectrogram. Prior to decoding, we concatenate four embeddings: phoneme encoder output, emotional vector, discrete SSL tokens, and a learned speaker embedding. These are projected into a unified representation, which is then linearly mapped to match the decoder's input dimension. This design ensures controllable, semantically aligned, and speaker-consistent synthesis.
\end{itemize}
\vspace{-2mm}

\subsection{Dataset Configuration}

To evaluate the performance of EmoSSLSphere under controlled conditions, we utilize single-speaker emotional corpora in English and Japanese. No speaker adaptation or cross-speaker fine-tuning is performed to isolate the effect of emotional and prosodic modeling. Additionally, for inference, we use ground-truth utterances as reference speech sampled from the same speaker and language as the target synthesis.

\begin{itemize}
    \item \textbf{English Dataset:} We use 80 utterances from the Emotional Speech Dataset (ESD)~\cite{zhou2022emotionalvoiceconversiontheory}, including 20 samples each for four emotion categories: Angry, Sad, Happy, and Surprised. All samples are from a single female speaker.

    \item \textbf{Japanese Dataset:} We use 60 utterances from the JVNV corpus~\cite{jvnv}, composed of 10 samples each for Angry, Sad, Happy, Surprised, Disgust, and Fear. The data is from a single female speaker, and evaluation samples are strictly separated from training data.
\end{itemize}

\vspace{-2mm}

\subsection{Evaluation Framework}

Our evaluation aims to verify whether EmoSSLSphere effectively enhances emotional expressiveness, intelligibility, prosody control, acoustic quality, and cross-lingual consistency. We compare EmoSSLSphere against two baseline methods to clearly isolate the contributions of our proposed components:

\begin{itemize} 
    \item \textbf{NATSpeech with Emotion Labels}:As a baseline, we implement NATSpeech~\cite{ren2022portaspeechportablehighqualitygenerative} trained separately on emotion-classified data subsets without explicit emotional embedding or style control. This represents a basic emotional synthesis system with limited controllability.
    
    \item \textbf{EmoSphere-TTS}: This model, as stated as 'conventional' model in some of future descriptions, employs only spherical AVD-based emotional vectors without integrating SSL-based discrete tokens or semantic text encodings. Comparing our method with EmoSphere-TTS~\cite{Cho_2024} helps demonstrate the added value of token-level prosodic modeling and semantic conditioning.
\end{itemize}

Additionally, to rigorously assess the individual contribution of each module in EmoSSLSphere, we conduct ablation studies by selectively removing or altering key components: 
\begin{itemize} 
    \item Reference waveform conditioning (removal and mismatch scenarios)
    \item Semantic text encoder (removal)
    \item SSL token discretization step (removal)
\end{itemize}

Evaluations encompass both objective metrics (e.g., Error Rate, SpeechBLEU, MCD) and subjective human ratings (Mean Opinion Scores—MOS), providing a comprehensive view of performance across linguistic, acoustic, emotional, and perceptual dimensions.

\section{Experimental Results}

We evaluate EmoSSLSphere across four dimensions: (1) Speech Intelligibility, (2) Acoustic Quality, (3) Naturalness, and (4) Emotional Expressiveness. Additionally, we conduct detailed analyses to understand the model’s behavior in multilingual emotional contexts.

\subsection{Speech Intelligibility Evaluation}\label{ssec:intel}

We first assess the intelligibility and semantic accuracy of synthesized speech. Evaluation metrics include \textbf{Error Rate (ER)}, \textbf{SpeechBERTScore}, \textbf{SpeechBLEU}, and \textbf{SpeechTokenDistance}\cite{saeki2024speechbertscorereferenceawareautomaticevaluation}. Error rates are measured by transcribing synthesized speech with the Whisper-base ASR model\cite{radford2022whisper} and comparing results against ground-truth transcripts using Levenshtein distance. Word Error Rate (WER) and Character Error Rate (CER) are utilized for English and Japanese, respectively.

Table 1 summarizes the results. EmoSSLSphere demonstrates superior performance compared to the baseline methods across most metrics, particularly in lower Error Rates and higher SpeechBLEU scores. This indicates enhanced semantic consistency and fidelity. Additionally, EmoSSLSphere consistently achieves strong intelligibility across both languages, highlighting the robustness of integrating semantic and prosodic representations via discrete SSL tokens and text conditioning.

\subsection{Acoustic Quality Evaluation}\label{ssec:quality}

Acoustic quality is assessed using objective metrics: \textbf{Mel Cepstral Distortion (MCD)}, \textbf{LogF0RMSE}, and \textbf{WARP-Q}~\cite{Wissam_IET_Signal_Process2022}. These metrics capture spectral distortion, pitch accuracy, and overall codec-based acoustic quality, respectively.

As presented in Table 2, EmoSSLSphere consistently outperforms baseline models in acoustic accuracy, indicated by significantly lower MCD values and reduced pitch estimation errors (LogF0RMSE). Improved WARP-Q scores also demonstrate enhanced acoustic fidelity, suggesting that EmoSSLSphere effectively preserves spectral and prosodic nuances necessary for realistic emotional expression.

\subsection{Naturalness Evaluation}

To measure naturalness, we employ the network-driven subjective metric \textbf{UTMOSv2}~\cite{baba2024utmosv2} along with human-based subjective Mean Opinion Scores (\textbf{nMOS}). Human evaluators (9 English and 21 Japanese native speakers) rated naturalness of synthesized speech samples from GT, baseline method, and EmoSSLSphere on a 1–5 MOS scale.

Table 2 shows that EmoSSLSphere achieves the highest naturalness ratings in both languages, confirming that the proposed method synthesizes speech that closely approximates human-like prosody and natural emotional expressiveness. The results clearly indicate that combining discrete SSL features with spherical emotion vectors contributes significantly to perceived naturalness.

\subsection{Evaluation of Emotional Expressiveness}

\begin{figure}[tb]
    \centering
    \includegraphics[width=7.9cm]{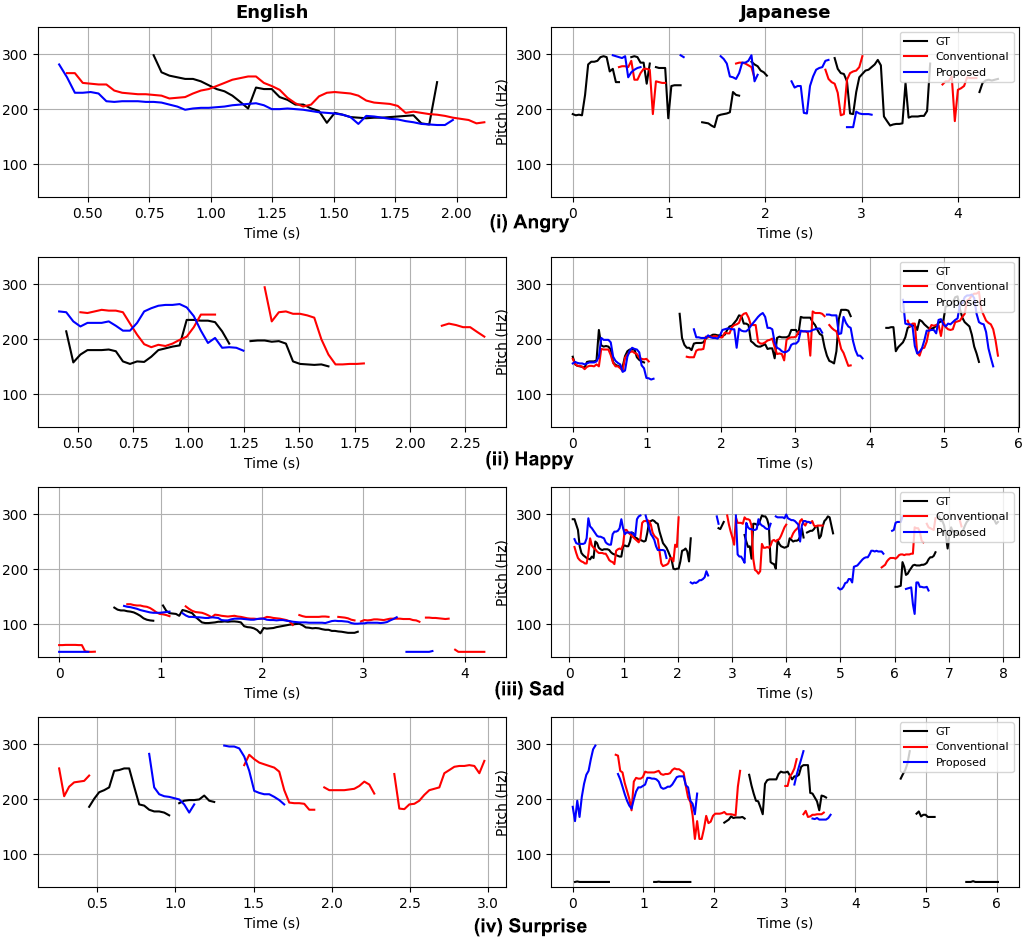}
    \caption{Pitch tracks of speech generated for different emotions (Angry, Happy, Sad, and Surprise) for each language}
    \vspace{-2mm}
    \label{fig:fig/pitch_tracksv2.png}
\end{figure}

To directly assess emotional expressiveness, we evaluate synthesized speech using objective metrics derived from Arousal-Valence-Dominance (AVD) vectors. Specifically, we compute the root mean squared error (RMSE) between the AVD values of synthesized speech and ground-truth samples, averaged per emotion category. Lower AVD RMSE values indicate that synthesized speech more closely matches the intended emotional expression.

Table 3 shows that EmoSSLSphere achieves the lowest average RMSE values for both languages, confirming that integrating discrete SSL tokens with spherical emotion vectors enhances fine-grained emotional control compared to baseline models. To further illustrate emotional controllability, we perform a qualitative analysis through pitch contour comparisons of synthesized speech (Figure 4). By examining representative pitch trajectories for four emotions (Angry, Happy, Sad, and Surprise), we observe that EmoSSLSphere effectively captures characteristic prosodic variations. As baseline models, our method consistently yields smooth and coherent pitch transitions, indicative of improved emotional expressiveness and natural prosodic flow.

\vspace{-1mm}
\subsection{Formant Analysis and Cross-Lingual Behavior}

To further investigate emotional articulation across languages, we conducted a formant analysis, examining variations in the first three formants (F1, F2, F3) of synthesized speech for each emotion (Figure 4). Formants serve as key acoustic indicators related to vowel articulation and vocal tract resonances.

Firstly, F1 reflects vowel height and mouth openness. Results indicate that EmoSSLSphere closely matches ground truth (GT), consistently producing higher F1 values for high-arousal emotions and lower values for lower-arousal emotions across both languages. Compared to EmoSphere-TTS, our proposed method more accurately captures nuanced emotional articulations, suggesting improved vowel quality control.

Secondly, F2 correlates with tongue frontness and vowel positioning. EmoSSLSphere maintains F2 trajectories that closely align with GT speech, particularly in Japanese, where vowel fronting differences are less pronounced compared to English. This consistency demonstrates EmoSSLSphere’s effectiveness in precise prosodic and emotional modeling across distinct linguistic contexts.

Lastly, F3 is associated with voice quality and timbre, affecting resonance and speaker characteristics. EmoSSLSphere effectively reproduces F3 patterns similar to GT, maintaining clearer timbral distinctions for different emotions—such as higher resonance for high-arousal emotions and subdued timbre for low-arousal emotions. This performance surpasses the conventional model, particularly in Japanese, where incorrect resonance shifts can significantly degrade naturalness.

Overall, these formant analyses confirm that EmoSSLSphere successfully generalizes emotional articulations across languages, maintaining phonetic and prosodic integrity more effectively than prior approaches. ~\cite{formant1, formant2}.

\begin{figure}[tb]
    \centering
    \includegraphics[width=7.9cm]{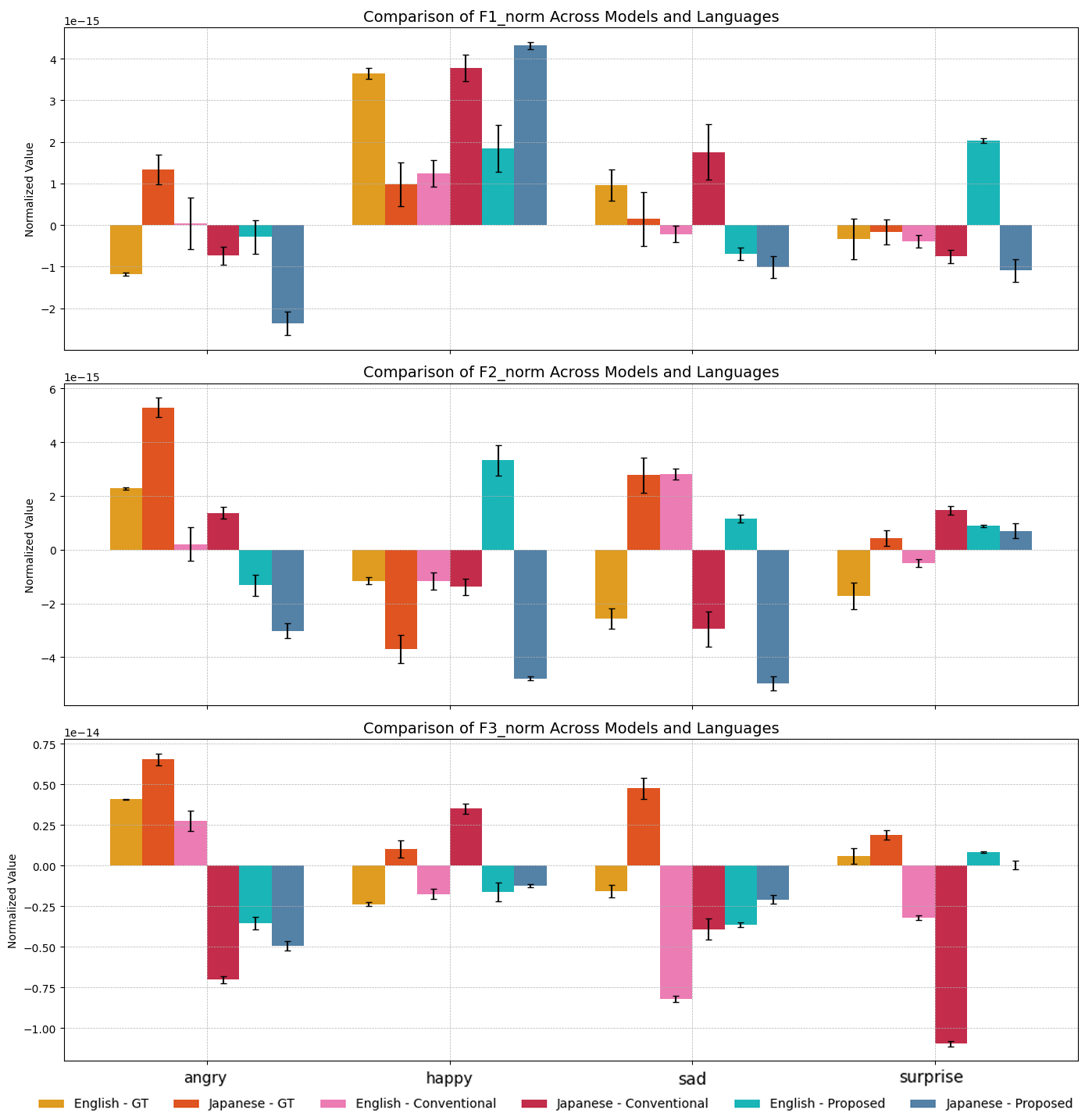}
    \caption{Analysis on formant variations (F1, F2, F3) of speech sample for each emotion and language}
    \vspace{-2mm}
    \label{fig:fig/pitch_analysisv2.png}
\end{figure}

\vspace{-1mm} \subsection{Ablation Studies}

To examine the contributions of individual components within EmoSSLSphere, we perform comprehensive ablations on key modules: (1) Reference waveform, (2) Semantic Text Encoder, and (3) k-means discretization of SSL features.

\begin{itemize} 
    \item \textbf{Without Reference Waveform:} Removing reference waveform conditioning not only degrades intelligibility but also significantly impacts acoustic quality and prosody. This confirms that exemplar-based conditioning via SSL-derived tokens strongly contributes to prosodic fidelity and emotional precision. Nonetheless, intelligibility remains relatively stable due to robust phoneme-level conditioning.
    
    \item \textbf{Reference Waveform with Unmatched Conditions:} To evaluate sensitivity to reference inputs, we tested mismatches in reference content, emotion, and speaker. Content mismatch notably reduced intelligibility and acoustic quality, while emotion mismatch moderately degraded emotional accuracy. Speaker mismatch had minimal adverse effects, which even exceeded proposed experiments in some of the intelligibility-sided metrics, suggesting robust speaker-independent prosody modeling. These results highlight the critical role of semantic and emotional alignment between reference and target utterances.

    \item \textbf{Without Semantic Text Encoder:} Excluding the semantic text encoder, while retaining the phoneme encoder, led to slight declines in prosodic and emotional naturalness. This indicates that high-level linguistic features from the semantic encoder significantly enhance emotional and prosodic consistency, even though they do not directly serve as decoder inputs.

    \item \textbf{Without k-means Discretization:} Removing discrete SSL tokens in favor of continuous HuBERT features led to noticeable performance degradation in intelligibility and acoustic metrics. The structured representation from discrete tokens evidently provides more stable prosodic and segmental alignment, supporting higher-quality synthesis.

\end{itemize}

Collectively, these ablation studies confirm the integral role of each module, validating our unified approach for multilingual emotional TTS synthesis.

\vspace{-1mm} \section{Conclusion}

This study introduced \textbf{EmoSSLSphere}, a multilingual emotional TTS framework that integrates spherical emotion vectors with discrete token features derived from SSL. By combining interpretable emotion control with prosody-aware token representations, EmoSSLSphere addresses key limitations in emotional TTS synthesis—particularly in emotional controllability and fine-grained prosody modeling.

Comprehensive evaluations demonstrated that EmoSSLSphere outperforms existing methods in terms of intelligibility, acoustic quality, naturalness, and emotional expressiveness in both languages. Notably, our approach enhances emotional articulation and pitch control, aligning closely with human speech in objective metrics. Ablation studies further confirmed the effectiveness of each component, especially the integration of discrete prosodic tokens and semantic conditioning.

Despite these promising results, several limitations remain. Current evaluations were conducted using single-speaker emotional datasets and limited-scale subjective assessments. Additionally, our study did not explicitly evaluate speaker fidelity or perceptual consistency across languages, and emotional authenticity was assessed via proxy metrics (e.g., AVD RMSE) rather than direct human judgments.

Future work will address these limitations by:
\begin{itemize}
    \item Expanding experiments to multilingual and multi-speaker settings to evaluate generalization across diverse speakers and languages.
    \item Conducting perceptual studies to evaluate speaker similarity, emotional authenticity, and cross-lingual prosodic transfer through human evaluations.
    \item Integrating EmoSSLSphere into semi-supervised architectures (e.g., EmoSphere++) to enhance training scalability while preserving prosodic control via discrete tokens.
\end{itemize}

By addressing these directions, we aim to further advance emotional and multilingual TTS, contributing toward more expressive, controllable, and robust speech synthesis technologies.

\newpage
\bibliographystyle{IEEEtran}
\bibliography{park}

\end{document}